\documentclass[letter,oldversion]{aa} 
\usepackage{graphicx}
\usepackage{txfonts}
\begin{document}
   \title{Spatially resolved submm imaging of the HR 8799
   debris disk}

   \author{J. Patience\inst{1}, J. Bulger\inst{1}, R. R. King\inst{1},
     B. Ayliffe\inst{1}, M. R. Bate\inst{1},
          I. Song\inst{2}, C. Pinte\inst{3},  J. Koda\inst{4},
          C. D. Dowell\inst{5},
          \and
          A. Kov{\'a}cs\inst{6}\fnmsep\thanks{Based on observations
          obtained at the Caltech Submillimeter Observatory}
          }

     \institute{Astrophysics Group, College of Engineering, Mathematics and Physical Sciences,
     University of Exeter, Exeter, EX4
              4QL, UK
	      \and
	      Department of Physics and Astronomy, University of
              Georgia, Athens, GA 30602-2451, USA
	      \and
	      UJF-Grenoble 1 / CNRS-INSU, Institut de Plan\'{e}tologie et d'Astrophysique de Grenoble UMR 5274, Grenoble, F-38041, France
	      \and
              Department of Physics and Astronomy, Stony Brook
              University, Stony Brook, NY 11794-3800, USA 
             \and
	      Department of Physics and Astronomy, Caltech, MC 249-17,
             Pasadena, CA 91125, USA
	     \and
	     School of Physics \& Astronomy, Tate Lab Room 148, 116
              Church Street S.E., Minneapolis, MN 55455, USA 
             }

   \date{Received June 03, 2011; accepted June 21, 2011}

 
    \abstract
   {Dynamical interactions between planets and debris disks may
     sculpt the disk structure and impact planetary
     orbits, but only a few systems with both imaged planets and
     spatially resolved debris disks are known.   
With the Caltech Submm Observatory (CSO), we have
observed the HR 8799 debris disk at 350$\mu$m. The 350$\mu$m
map is the first spatially resolved measurement of the debris disk
encircling the HR 8799 planetary 
system at this wavelength. Both the flux and size of the
emission are consistent 
with a Kuiper belt of dust
extending from $\sim$100-300 AU. Although the resolution of the current map
is limited, the map shows an indication of offset asymmetric
  emission, and several
scenarios for this possibility are explored with radiative
  transfer calculations of a star-disk system and N-body
numerical simulations of planet-disk interactions with parameters
  representative of the HR 8799 system. 
  }


   \authorrunning{Patience, Bulger, King, Ayliffe, Song, Pinte, Koda,
     Dowell, \& Kovacs}
   \titlerunning{Submm imaging of the HR 8799 disk}

   \maketitle
%

\section{Introduction}
Planets and debris disks are intimately linked, as planets can scatter
or trap planetesimals in a disk (Mouillet et al. 1997), and planetestimals may
drive planet migration (Kirsh et al. 2009). Three planetary systems have been
imaged orbiting dusty 
A-stars (Marois et al. 2008; Kalas et al. 2008; Lagrange et al. 2010) with remnant debris disks (Smith et al. 1984; Holland et al. 1998; Zuckerman \& Song 2004) sustained by the
collisional grinding of planetesimals into smaller particles (Backman \& Paresce 1993).  These
systems present rare cases in which it is possible to
explore the dynamical effects on disk structure of planets with known
locations. Structures such as asymmetries and clumps can encode the
effects of gravitational interactions (e.g. Liou \& Zook 1999; Kuchner \& Holman 2003; Wyatt 2006; Quillen \& Faber 2006). 
The submm/mm wavelength range is ideal for studying the interaction of
the disk and planets, since the large
grains are expected to remain in resonances with the planets, while
the smaller grains evolve into axisymmetric structures due to
scattering or radiation pressure (Wyatt 2006).

From {\it IRAS} photometry, HR 8799 was identified as a debris disk
system with excess emission above the stellar photosphere at 60$\mu$m,
consistent with cool dust (Zuckerman \& Song 2004). With sensitive
{\it Spitzer} measurements over the 5.5$\mu$m-35$\mu$m range, the disk
was further estimated to contain two components: a hotter inner ring
of dust analogous to the asteroid belt with an
inner radius of 6 AU and a cooler outer disk (Chen et al. 2006). The presence of
dust, combined with a young age estimated from the colour-magnitude
diagram, made HR 8799 a target for high-contrast imaging which
revealed a system of four giant planets at projected separations
ranging from $\sim$14--70 AU (Marois et al. 2008, 2010). The
planets are located between the two dust
populations. Subsequent {\it Spitzer} images at 70$\mu$m resolved a
distribution of dust at very large radii from HR 8799, and this
emission may probe smaller blown out grains (Su et al. 2009). This letter 
reports the first observations of the disk
at 350$\mu$m, including a spatially
resolved map and a comparison with numerical models.

\section{Observations}
Submillimeter observations of HR 8799 were acquired with the 10m Caltech
Submillimeter Observatory (CSO) with the SHARCII instrument
(Dowell et al. 2003) which includes a 12 $\times$ 32 bolometer array with a
$>$90\% filling factor. At the observation wavelength
of 350 $\mu$m, the beam size is $\sim$8$\farcs$5 and the shape is
stabilised by the Dish Surface Optimization System (DSOS) which
corrects for dish figure changes as a function of telescope
elevation. The Lissajous 
scanning pattern employed for the observations resulted in a map size of
$\sim$3$\farcm$3$\times$2$\farcm$0. 
The data were acquired
over two nights with  excellent tau conditions
($\tau_{225GHz}$ = 0.03-0.04) on 6 \& 7 Aug 2007. Follow-up telescope
time allocated to obtain deeper maps did not 
have sufficient tau conditions. For the observing 
sequence, a series of 4-6 scans of 300s to 620s each were obtained on the
target and were preceded and followed by 120s scans of a bright
pointing and flux 
calibrator, Uranus, Neptune, or the quasar 3C454.3. In total, 20
scans were taken on the science target for a total of 2.75 hours of
data. Given the importance of calibration, a similar number of 23
scans of calibrators were recorded over the time period of the
observations. The target and calibrators sampled a similar range of
temperature, elevation, and opacity. 


\section{Data analysis}
The construction of the CSO map required two main steps -- calculating
pointing offsets and extracting the signal from the background. For
each scan, a pointing correction was first calculated by first
computing and then applying pointing corrections that account for both
static and time variable pointing drifts, using calibrator scans, a pointing model\footnote{www.submm.caltech.edu/sharc/analysis/pmodel/.} and
software. The calibration scans were used in conjunction with the
pointing model to calculate the pointing offset applicable at the time
and sky position of each target observation. The pointing corrections,
along with the measurements of the atmospheric opacity at the time of
observation, were then included in the data processing using the
software CRUSH, version 2.01-4 (Kov\'{a}cs 2008). In the final map, the total flux was
measured within the 3$\sigma$ contour using the MIRIAD  (Sault et al. 1995)
software. The absolute flux calibration was based on the scans of the
planet Uranus and the planet flux at the observation date from the JPL
Horizons model. The total uncertainty in the HR 8799 flux is 35\%,
a combination of the small (6\%) variation in the Uranus
flux measurements and the rms noise level in measured in the final
source map. 

To quantify the location of the emission peak, the distance from the
host star position to the 
brightest part of the disk was measured. Since the
photosphere of HR 8799 is not detectable at 350$\mu$m, it was necessary to
identify the location of the central source based on the telescope
pointing. The bright calibration targets were used to empirically
measure the pointing accuracy. For the pointing test, each calibrator
scan was reduced in a manner analogous to the targets - i.e. by
removing that calibrator scan (and others taken within a few minutes)
from the pointing correction calculation and deriving a new pointing
correction from the remaining calibrators and pointing model and then
applying the pointing offset to the CRUSH reduction. This approach
should be conservative in calculating the pointing offset to apply to
the calibrator scan since the calibrators for the target are taken at
most 30min before/after the targets, while this approach for the
calibrators has mainly included calibrators $\sim$60min before/after each
calibrator scan treated as a pointing test. The absolute value of the
pointing error for each calibrator analysed with this procedure was
measured from the difference of the position of the calibrator in each
map to its known coordinates. Since we are interested in the central
position of the source, we calculated the standard deviation of the
mean of the absolute offsets. Based on this empirical test, the
position uncertainty is $\pm$0$\farcs$6. Since the position of
the emission peak is identified within a fraction of a pixel in the
final map,  the source position uncertainty
dominates the uncertainty on distance measurements.


\section{Numerical modelling}
To interpret the debris disk map, the flux and morphology were
  compared with three numerical models -- a radiative transfer model
  of a star surrounded by symmetric zones of dust, a simulation of
  massive planets migrating outwards and interacting with
  planetesimals and dust, and a simulation from the literature of a
  low mass planet interacting with planetesimals. The Monte Carlo 3D continuum radiative transfer code MCFOST
(Pinte et al. 2006) was used to generate SED models and images at 350$\mu$m.  In the MCFOST routines, the
photons from the central star with properties given in Reidemeister et al. (2009) were propagated through the disk with
a model incorporating a combination of Mie theory scattering,
absorption, and re-emission. The disk parameters used to construct the SED model are given in
Table 1. The disk zones are based
on the 3-component model described in Su (2009). Minor
modifications were
included to account for the dynamically cleared chaotic zones around
the innermost and outermost planet
  (Quillen \& Faber 2006; Moro-Mart\'{i}n et al. 2010; Fabrycky \& Murray-Clay 2010). A series
  of simulated images were generated for a range of outer disk radii
  and inclinations, and Table 1 shows the values most consistent with
  the CSO map.

The structure in the CSO map was also compared with numerical models
performed using an N-body code. Our numerical model included the
interaction of planets resembling HR 8799b (6 M$_\mathrm{Jup}$) (Currie et al. 2011) and HR 8799c
(8 M$_\mathrm{Jup}$) (Currie et al. 2011) migrating outwards at the same rate through a disk of
planetesimals. Since there was no significant difference in the
  results when two planets were included rather than one, the planets more
  distant planets from the debris disk should have no impact and were not added to the
  simulation. The initial planetesimal belt had a width of 5 AU and 
an inner radius of 91 AU, and the planets migrated 15 AU to their
final orbits. Once the two planets had reached approximately their
present day orbits (Marois et al. 2008), the orbits were circularised, and the
planetesimal disk replaced by a dust disk. The stability of the
planetesimals in a very similar configuration has already been shown
in previous simulations (Moro-Mart\'{i}n et al. 2010).

The transformation of
planetesimals to dust entailed the introduction of radiation pressure
and Poynting-Robertson drag through the parameter $\beta$ that quantifies
the ratio of radiation to gravitational forces (Burns et al. 1979). The grains
responsible for the 350$\mu$m emission have a size distribution peaked
near 350$\mu$m, so the $\beta$ value appropriate for the map is
0.0055. The model is designed to investigate the inner region of the
Kuiper belt and does not include the entire disk structure
or all three components of the SED
model.
Simulated surface density maps of the inner Kuiper belt were produced
at a range of wavelengths. Finally, the CSO map structure was also
  considered in the context of results of dynamical
  simulations of a low mass planet interacting with a planetesimal
  belt (Reche et al. 2008).


\section{Results and discussion}
\subsection{Disk flux and size}
The CSO map is given in Figure 1, and the measured flux for the
disk integrated over the 3$\sigma$ contour is 89 $\pm$ 26 mJy. The 350$\mu$m map of the HR 8799 debris disk reveals emission that is
extended compared to the maps of the bright point source calibrators
taken on the same night. Only spatially resolved images of dust with small $\beta$
values are capable of distinguishing the location of the dust based on
the structure, since the SED solution for the disk radius is
degenerate with dust properties such as a$_{min}$. The inner disk is
not resolved, but set at 100 AU, based on previous simulations of
orbital stability (Moro-Mart\'{i}n et al. 2010) and the mass of the
outermost planet.

To estimate the size of the dust belt, a series of MCFOST models were run with a
range of outer disk radii. All models simultaneously matched the 350$\mu$m flux 
and previously measured fluxes reported in Su et al. (2009)
or compiled by Reidemeister et al. (2009), plotted in Figure 2. Although the map resolution
and sensitivity are limited, the disk outer radius is consistent with
$\sim$300 AU, and this radius is marked on Figure 1, along with the
assumed inner radius. The 350 $\mu$m emission is largely confined
within a radius of 300 AU. Further details on the set of simulated
350$\mu$m images from MCFOST are given in Figure 3. A radius of 200
AU results in a more compact structure inconsistent with the data, though higher sensitivity maps will
be required to rule out disks much larger than 300 AU, since the
surface brightness of the most extended emission is beyond the dynamic
range of the CSO data. 

\begin{table}
\begin{minipage}[t]{\columnwidth}
\caption{MCFOST Disk Model Parameters}
\label{catalog}
\centering
\renewcommand{\footnoterule}{}  
\begin{tabular}{lccc}
\hline \hline          
Model  & Asteroid & Kuiper & Outer \\ 
Parameter \footnote{dust mass $M_{dust}$, 
inner radius $r_{in}$, outer radius $r_{out}$, scale height $H_0$ at a
reference radius r$_o$, flaring profile
exponent for the disk height $H(r)$,
surface density profile $\Sigma(r)$, minimum grain size $a_{min}$, maximum
grain size $a_{max}$ and the differential grain size distribution 
dn/da.} & Belt & Belt & Halo   \\ 
\hline                    
M$_{dust}$ [M$_{\sun}$] & 3.3 $\times$ 10$^{-12}$ & 3.6 $\times$
10$^{-7}$ & 5.7 $\times$ 10$^{-8}$   \\
r$_{in}$ [AU]  & 6 & 100 & 300   \\
r$_{out}$ [AU]  & 8 & 300 & 1000   \\
H$_o$(r$_{in}$) [AU]  & 0.6 & 9 &  30   \\
H(r)  & $\sim$r${^0}$ & $\sim$r${^0}$ & $\sim$r${^0}$    \\
$\Sigma$(r)    & $\sim$r$^{0}$ & $\sim$r$^{0}$ &  $\sim$r$^{-1}$   \\
a$_{min}$ [$\mu$m]   & 2.0 & 3.0 & 1.0   \\
a$_{max}$ [$\mu$m]     & 4.5  & 1000 & 10   \\
dn/da   & $\sim$a$^{-3.5}$ & $\sim$a$^{-3.5}$ & $\sim$a$^{-3.5}$   \\
\hline
\end{tabular}
\end{minipage}
\end{table}

   \begin{figure*}
   \centering
\includegraphics[scale=0.2]{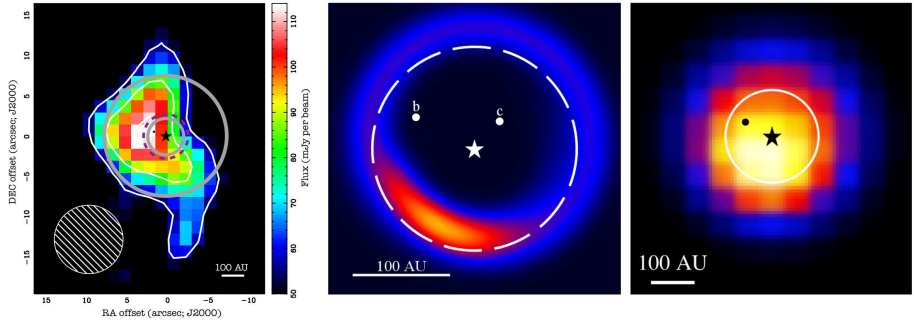}
      \caption{ {\bf(left)} The CSO 350$\mu$m image with the positions of the
        star (black star) and
        outermost planet (dot) indicated. The solid circles trace the
        locations of the inner and outer edge of the Kuiper belt used
        in the SED model and the dashed line shows the expected
        position of the 2:1 resonance.  {\bf (middle)} The disk
        surface density map from the simulation.
        The semi-major axis centers around the 2:1 resonance with the
        outer planet (dashed line), 
        and the moderate eccentricities ($\sim$0.3) of the dust grains explain
        their displacement from the circular path. {\bf(right)} The
        same numerical simulation convolved with an 
        8$\farcs$5 Gaussian to compare with the brightness peak
        position in the CSO map 
        of the same resolution. The 2:1 resonance is indicated by the white line.
              }
         \label{wyatt}
   \end{figure*}

   \begin{figure}
   \centering
   \includegraphics[scale=0.55]{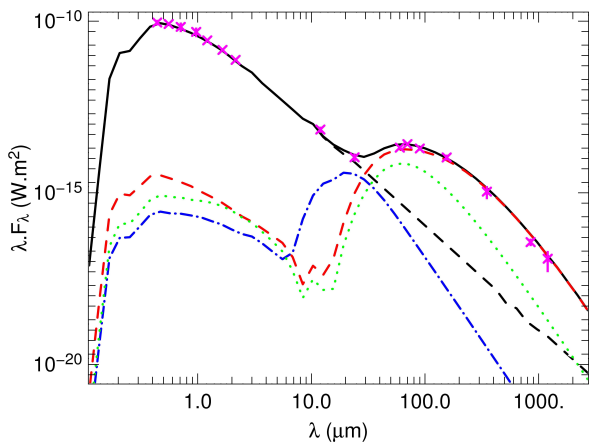}
      \caption{The fluxes of the HR 8799 star/disk system are plotted as
        a function of wavelength  (purple crosses), along with the SED model with the
        parameters listed in Table 1 (solid black line). Contributions
        from the stellar flux (black
        dashed line) and scattered and emitted light from the asteroid belt (blue
        dash-dot line), the Kuiper belt (dashed red line), and outer
        halo (green dotted line).
              }
         \label{sed}
   \end{figure}

\subsection{Disk structure}

The observed disk morphology is compared with two types of models of
the disk structure -- a symmetric dust distribution simulated with
MCFOST and an asymmetric dust distribution governed by gravitational
interactions and radiation pressure. Assuming a symmetric
circular ring for the disk described in Table 1, the effect of
changing the inclination is explored in Figure 3. As the emission
is optically thin at 350$\mu$m (unlike scattered light images at
shorter wavelengths), the disk pattern remains symmetric as the
inclination increases. Since the observed map is not well matched by
these symmetric patterns, it is not possible to place a strong
constraint on the disk inclination to compare with estimates from
planet astrometry (Lafreni\`{e}re et al. 2009), 70$\mu$m emission
(Su et al. 2009), and stability (Moro-Mart\'{i}n et al. 2010).

The CSO map is also compared with numerical models resulting in an
asymmetric dust distribution. 
In addition to the extended structure, the HR 8799 disk emission at
350$\mu$m is slightly offset from the coordinates of the host star. 
The distance from the star to the brightest point in the disk is
2$\farcs$1 $\pm$ 0$\farcs$7, or 84 $\pm$ 28AU. While the
uncertainty is too large to assign a specific 
value to the radius of the brightest region of the disk, an estimate can be
compared with the distance to the peak in surface density of the
numerical simulation in Figure 1. A dominant one-sided arc of higher surface density describes the
distribution of grains with $\beta$ values on the order of 0.005 or less, as
they become trapped in the 2:1 mean motion resonance (Wyatt 2006), based on
previously published simulations of migrating planets with masses up
to 1 M$_\mathrm{Jup}$ (Wyatt 2006). Our
simulation of a planetesimal belt is a test case 
extending to include two higher mass planets; a similar asymmetric pattern
is evident in our dust surface density map in Figure 1. The CSO map of the debris
disk shows a structure suggestive of a single bright clump like the pattern
produced by dust grains trapped in the 2:1 mean motion resonance with
a planet, while the map does not reveal two distinct clumps separated
by 180 degrees, as expected for material trapped in the 3:2
resonance (Wyatt 2006). Within the uncertainty of the measurement, the radius of
the brightest region of the disk is consistent with the peak in the
surface density image in our simulation; a convolution of the
simulated map with the CSO beam showing a similar offset is given in Figure 1. 

Based on previous simulations, resonant trapping cannot easily explain
that the brightest clump leads rather than lags the planet position,
since the libration point lagging position has a higher probability to
be filled for an outward migration with a planet with eccentricity
greater than 0.03 (Wyatt 2006). In our simulation of a planetesimal belt
surrounding two massive planets with slow outward migration, the
azimuthal angle of the bulge is subject to large scale oscillations
over the course of several orbits, moving between a leading and a
trailing position in a continuous manner, without crossing in front of
the planet. Such an oscillating structure can be produced if the
orbits of the planets become eccentric as they migrate (due to mutual
interactions) and, once formed, the structure can persist even if the
eccentricities of the planetary orbits are later damped. 

If dynamical interactions are governing the particle spatial
distribution such that the 350$\mu$m-emitting grains are trapped in a 2:1
mean motion resonance with the outermost planet, then there are
important implications for the orbital history of HR 8799b. To enhance
the population of planetesimals in the resonance, orbital migration is
required, although the rate of migration cannot be too high, or the
trapping probability will decrease (Reche et al. 2008). Planet migration may have
strongly influenced the timing and mass flux of the Late Heavy
Bombardment of the terrestrial planets in the Solar System (Gomes et al. 2005), so
evidence of orbital migration may be an important factor for the
conditions in the as-yet unexplored terrestrial planet region in the
HR 8799 system. Theoretical arguments favouring Currie et al. (2011) and rejecting Dodson-Robinson et al. (2009)
orbital migration for the HR 8799 system have also been proposed. For
a resonant pattern to persist, the eccentricity of the orbit of the
planet needs to be low, since the libration amplitude of the
planetesimals in resonance is increased as eccentricity increases,
causing the distribution to become smooth (Reche et al. 2008). Numerical simulations of
the HR 8799 planet system (prior to the discovery of the innermost
planet (Marois et al. 2010)), have identified a set of possible stable orbital solutions,
including two with low eccentricities for HR 8799b of 0.008 or 0.014 (Go\'{z}dziewski \& Migaszewski 2009). The clumped structure of the CSO map favours these low
eccentricity solutions. For comparison, the eccentricity of the
orbit of Neptune is 0.0112 and the migration of
Neptune may have caused the objects to move into the resonances (Malhotra 1993).  

An alternate, non-resonant mechanism for producing a single disk clump
is the presence of an outer, low mass planet on an eccentric orbit (Reche et al. 2008),
an interesting possibility for a system that already includes four
giant planets. In this case, the planet mass would be very low,
comparable to an Earth mass, the eccentricity would be large
($\sim$0.4-0.7) and the clump would appear at apastron. The lifetime for
non-resonant structures is expected to be lower, on the order of 35
Myr (Reche et al. 2008), though the recent assessment of the HR 8799 age of 30 Myr (Zuckerman et al. 2011) is
within the feature lifetime, and the system is old
enough to have passed from the runaway growth to the oligarchic growth
phase of terrestrial planet formation, with enough time for 
terrestrial planet formation (Chambers 2001).

\section{Future prospects}
The CSO map at 350$\mu$m has shown possible indications of dynamical
sculpting of the HR 8799 debris disk from the gravitational influence
of the outermost planet, and future multi-wavelength observations should reveal even
finer details. Simulations of the surface density distribution of
  dust particles emitting at 24$\mu$m, 70$\mu$m, 350$\mu$m, and
  850$\mu$m are given in Figure 4. At the shortest wavelength of
  24$\mu$m, a symmetric ring is expected and may be resolvable with
  8m-class telescopes, though below the limit of {\it Spitzer}. Both
  {\it Herschel} operating at 70$\mu$m and SCUBA2 at 450$\mu$m have
  resolution limits comparable to the CSO, and the
  70$\mu$m emission may be more distributed than the 350$\mu$m
  map. The ALMA and SMA interferometers operate at 450$\mu$m and
  850$\mu$ with higher positional accuracy than a single dish
  telescope, and the 850$\mu$m peak may be offset from the 450$\mu$m
  peak location. ALMA, in particular,  will have significantly higher sensitivity to improve the
measurement of the disk structure, and will be ideal to
pursue deeper, high fidelity maps for further
insights into the planet formation process in this benchmark
system.


\begin{acknowledgements}
       We gratefully acknowledge support to Exeter from the Leverhulme
       Trust (F/00144/BJ ) and STFC (ST/F0071241/1, ST/H002707/1) and
       to IPAG from EC's 7th Framework Program (PERG06-GA-2009-256513)
       and ANR of France (ANR-2010-JCJC-0504-01). We acknowledge
       collaborative support through the EC CONSTELLATION Network and
       helpful conversation with H. Beust and J-C
       Augereau. Observations at the Caltech Submillimeter Observatory
       have been supported by NSF grant AST 05-40882 and 08-38261. We
       thank the referee for helpful comments.

\end{acknowledgements}

\Online

\begin{figure*}
\centering
\includegraphics[scale=0.25]{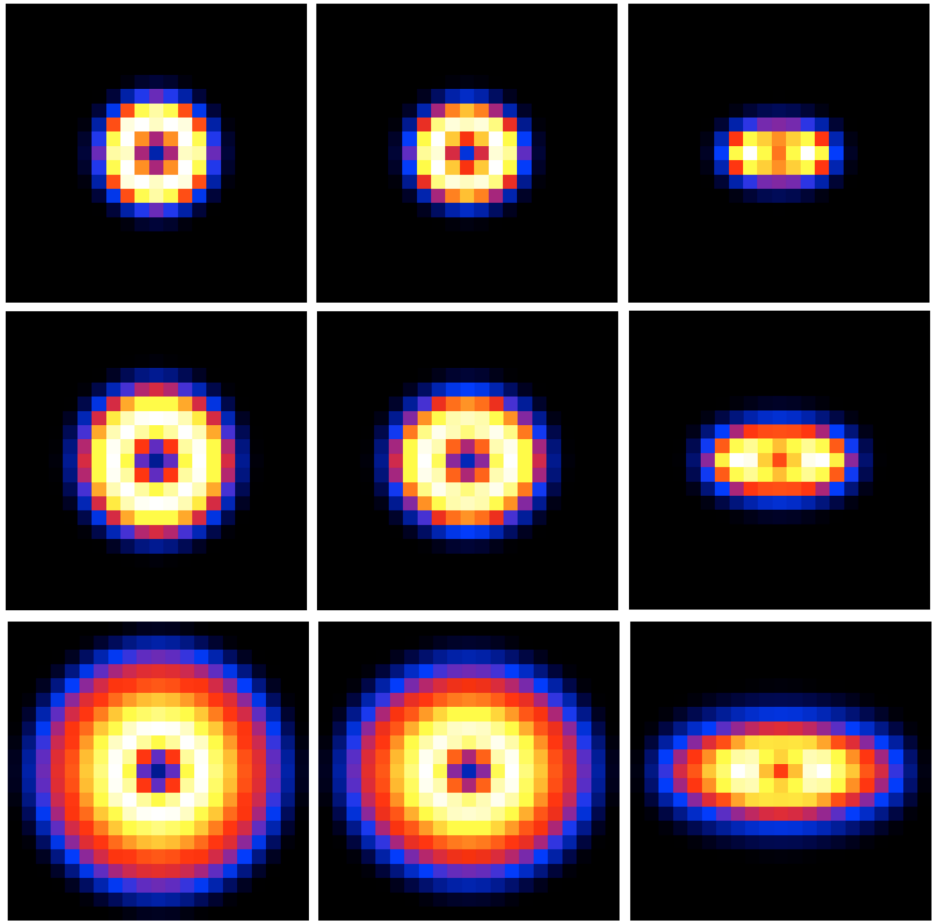}
      \caption{Model ray tracing images at a wavelength of 350$\mu$m generated with the MCFOST code
    for combinations of three outer disk sizes and disk
     inclinations. {\bf (top row, from left to right)} 200 AU outer disk with
      inclinations of 0$^{\circ}$, 30$^{\circ}$, and 60$^{\circ}$. {\bf (middle row)} 300 AU outer disk with
     same sequence of inclinations. {\bf (bottom row)} 500 AU outer disk with
    same sequence of inclinations. Each image has a size of 1200 AU
     $\times$ 1200 AU and the pixels match the size of the CSO
      map pixels. The dynamic range is a factor of 10.}
\label{pointdisk}
\end{figure*}

\begin{figure*}
\centering
\includegraphics[width=4cm]{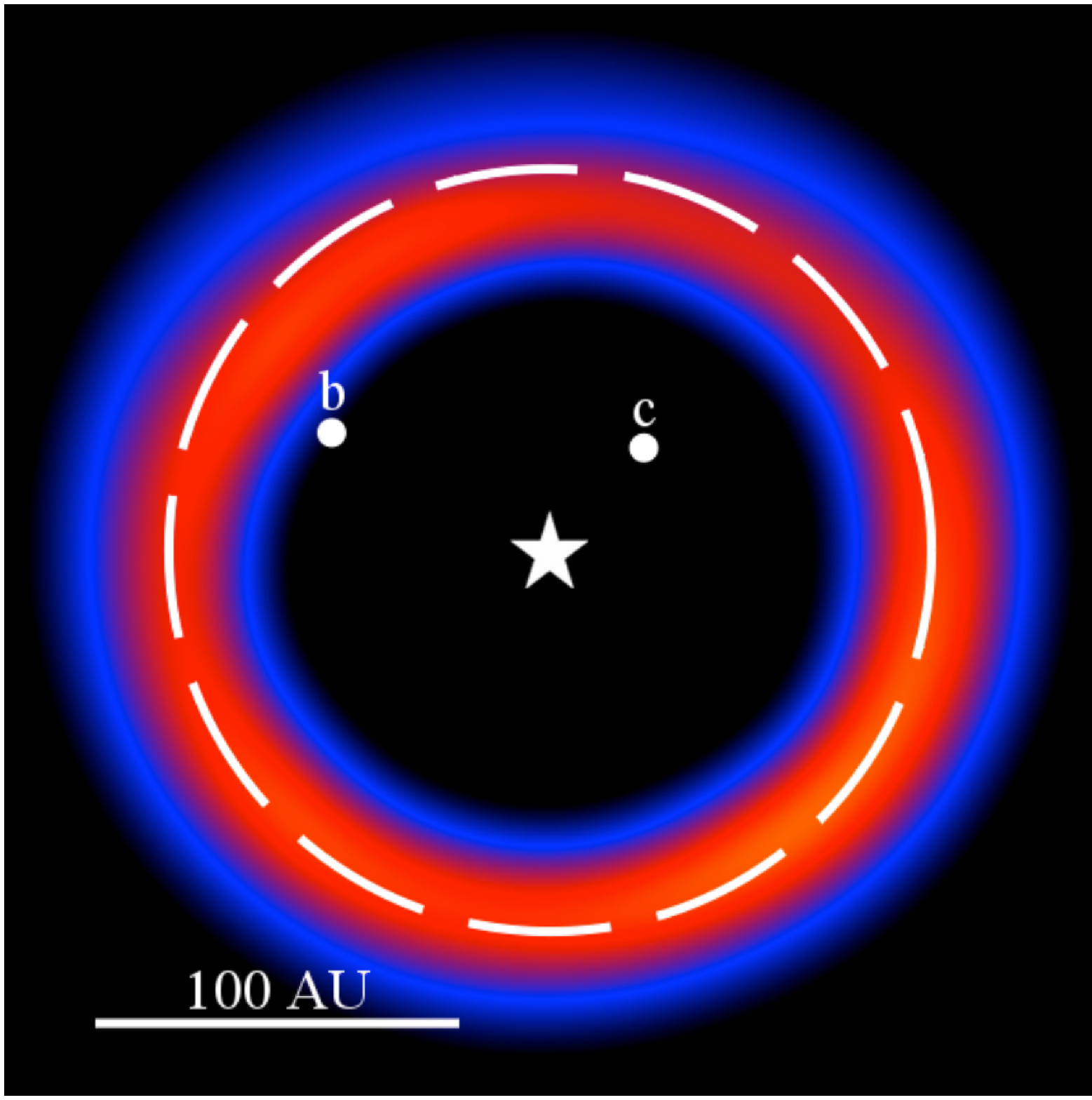}
\includegraphics[width=4cm]{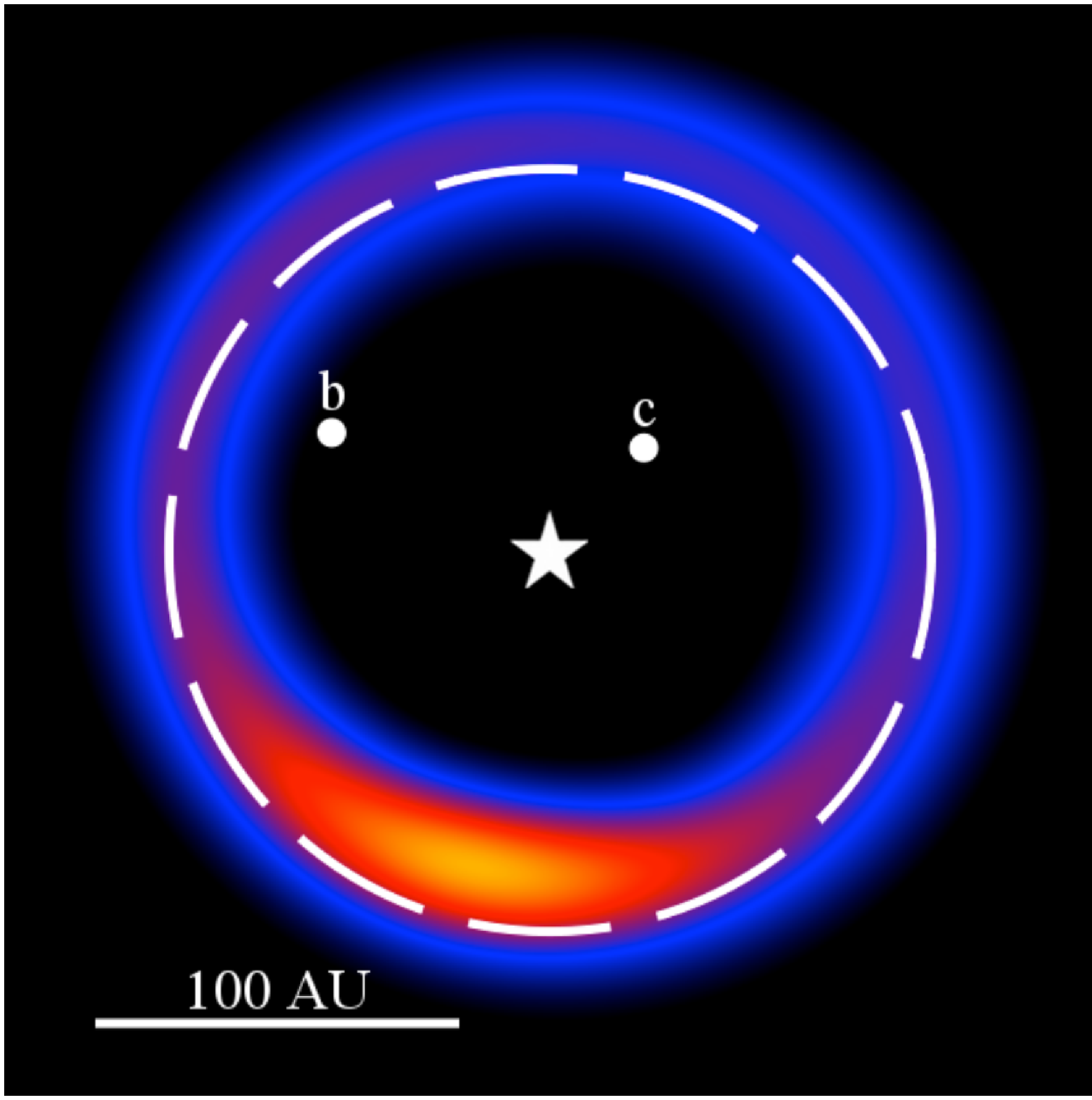}
\includegraphics[width=4cm]{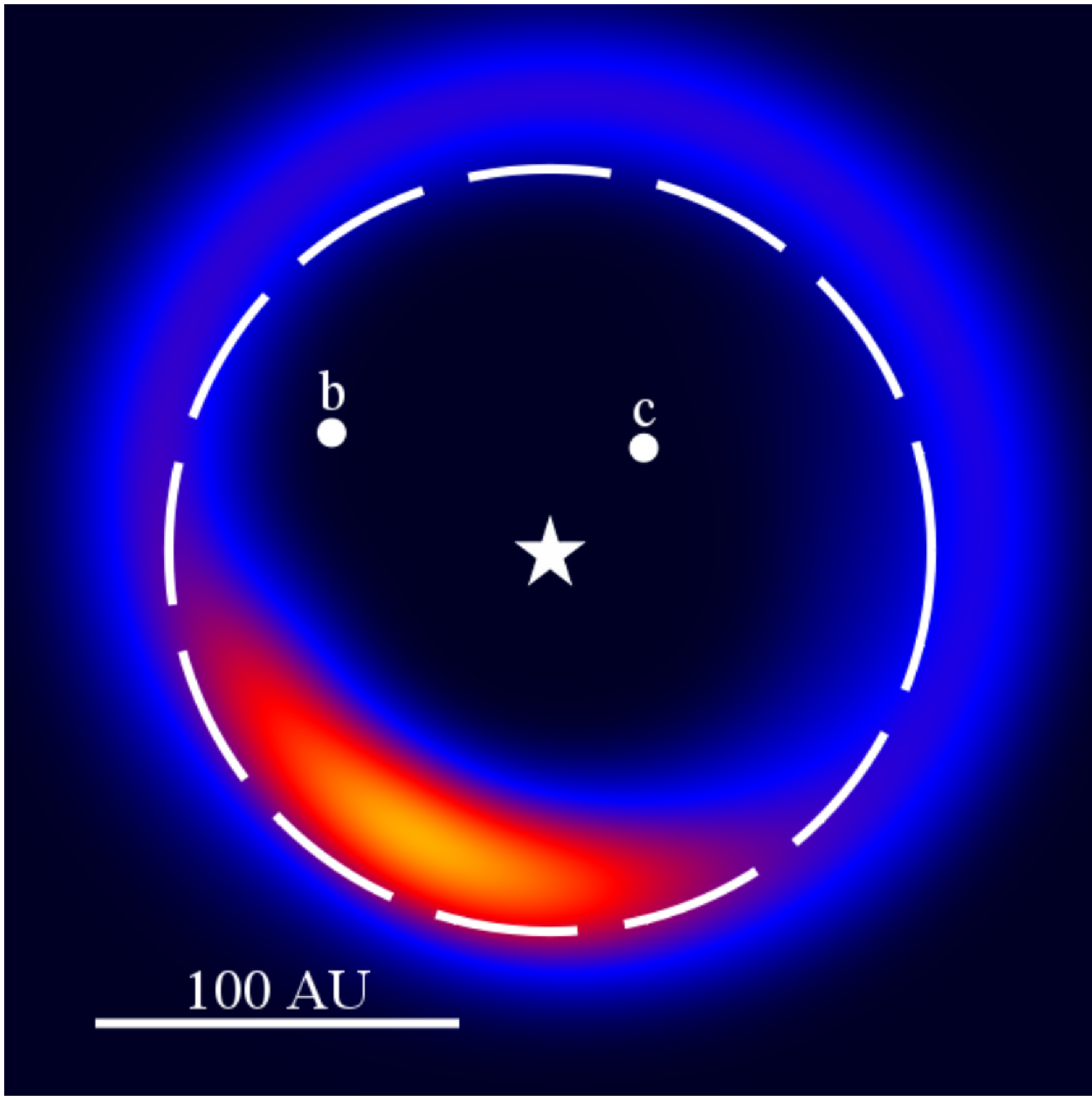}
\includegraphics[width=4cm]{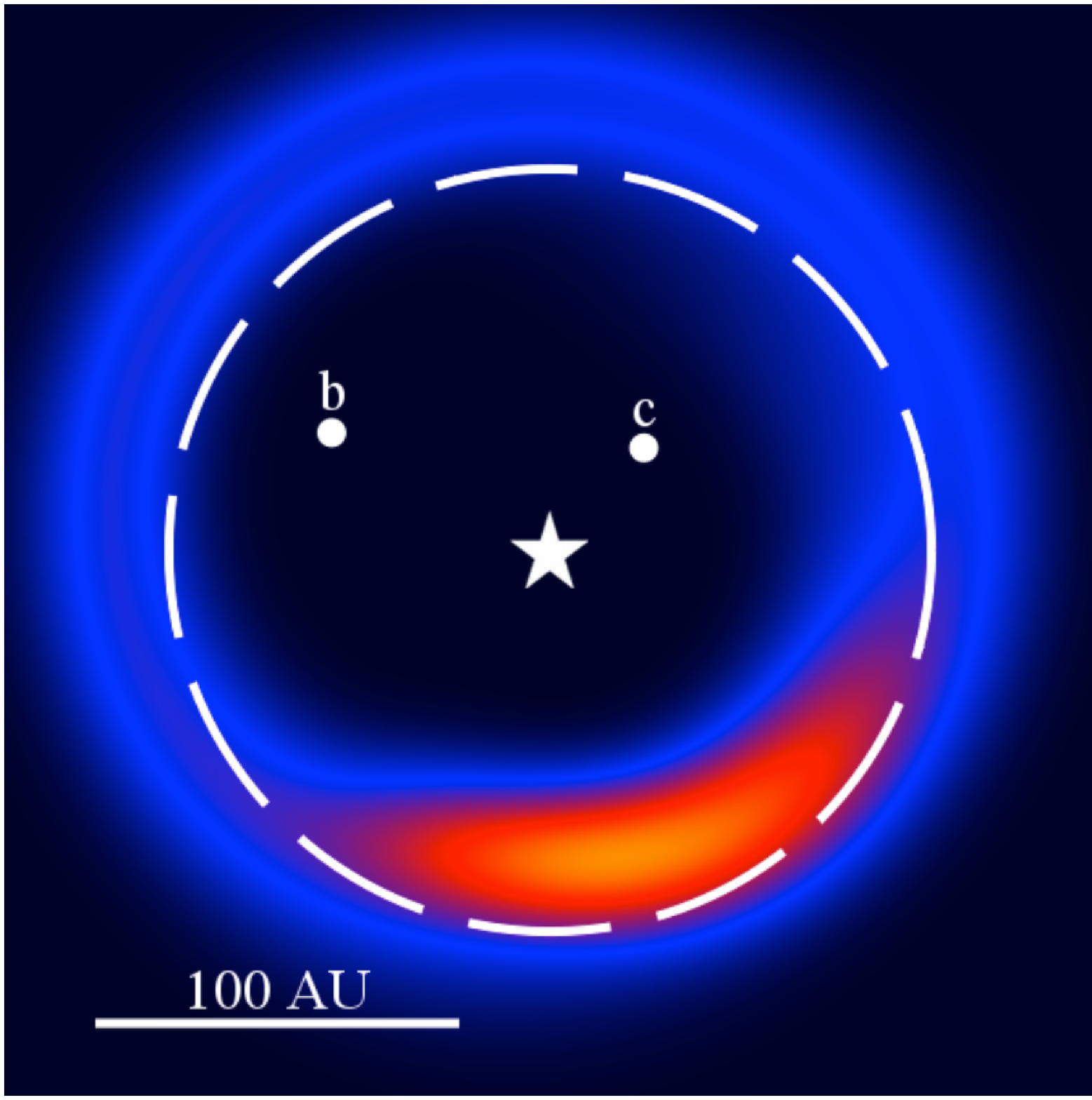}
\caption{Simulations of surface density from the N-body numerical model
 at four wavelengths, increasing from left to right: 24$\mu$m,
 70$\mu$m, 350$\mu$m, and 850$\mu$m. The dynamic range is a factor of
 10, and the simulation is
 smoothed to a resolution of 1$\farcs$5, matching the capability of
 ALMA. All figures are contemporaneous in the evolution.}
\label{appfig}
\end{figure*}

\end{document}